\documentclass{article}
\usepackage[utf8]{inputenc}
\pdfoutput=1

\usepackage{authblk}

\usepackage[backend=bibtex,sorting=none,url=false,isbn=false]{biblatex} 
\bibliography{main}

\usepackage[titletoc,title]{appendix}

\usepackage{mathtools} 
\usepackage{amsfonts}
\usepackage{dsfont}
\usepackage{bm}

\usepackage{algorithmic}
\usepackage[labelfont=bf]{caption}

\algsetup{linenodelimiter=\ }
\usepackage{algorithm} 
\usepackage[usenames,dvipsnames]{color}
\usepackage[colorlinks,citecolor=Violet]{hyperref}
\hypersetup{
    linkcolor = RedOrange,
    urlcolor = RoyalBlue
}

\usepackage[capitalise]{cleveref} 

\usepackage[normalem]{ulem}

\newcommand{\ind}{\mathds{1}}

\newcommand{\kn}{k_{(n)}}

\newcommand{\fkn}{f(\kn;\, N,K,n)}

\newcommand{\knstar}{k_{(n)}^*}

\newcommand{\kL}{k_{(L)}}

\newcommand{\kex}{k_{(5)}}

\newcommand{\pkw}{{p_{(k,w)}}}

\newcommand{\mukw}{{\mu_{(k,w)}}}
\newcommand{\mukwmk}{{\mu_{(k-1,w)}}}
\newcommand{\mukwmw}{{\mu_{(k,w-1)}}}

\newcommand{\hgp}{p^{\textrm{\tiny{HG}}}_{(n)}}
\newcommand{\hgpm}{p^{\textrm{\tiny{HG}}}_{(n-1)}}
\newcommand{\hgpex}{p^{\textrm{\tiny{HG}}}_{(5)}}
\newcommand{\hgpexa}{p^{\textrm{\tiny{HG}}}_{(4)}}
\newcommand{\hgpexb}{p^{\textrm{\tiny{HG}}}_{(6)}}

\newcommand{\vex}{\bm{v}_{\textrm{ex}}}
\newcommand{\vnull}{{\bm{v}^0}}
\newcommand{\vnullrnd}{{\bm{V}^0}}

\newcommand{\fk}{{f(k;\, N,K,n)}}
\newcommand{\fkdot}{{f(k;\, \dots)}}

\newcommand{\fkp}{{f(k+1;\, N,K,n)}}
\newcommand{\fkpn}{{f(k;\, N,K,n+1)}}
\newcommand{\fkpp}{{f(k+1;\, N,K,n+1)}}
\newcommand{\fkmm}{{f(k-1;\, N,K,n-1)}}
\newcommand{\fkmn}{{f(k;\, N,K,n-1)}}
\newcommand{\fkm}{{f(k-1;\, N,K,n)}}
\newcommand{\fn}{{f(n;\, N,K,n)}}
\newcommand{\fnmm}{{f(n-1;\, N,K,n-1)}}
\newcommand{\fK}{{f(K;\, N,K,n)}}
\newcommand{\fKmn}{{f(K;\, N,K,n-1)}}

\newcommand{\mkw}{{m_{(k,w)}}}
\newcommand{\mkwstart}{{m_{(0,0)}}}
\newcommand{\mkwmk}{{m_{(k-1,w)}}}
\newcommand{\mkwmw}{{m_{(k,w-1)}}}

\newcommand{\ckw}{{c_{(k,w)}}}
\newcommand{\ckwmk}{{c_{(k-1,w)}}}
\newcommand{\ckwmw}{{c_{(k,w-1)}}}

\newcommand{\mhgstat}{{s^{\textrm{\tiny{mHG}}}}}

\newcommand{\mhgstatnullrnd}{{S^{\textrm{\tiny{mHG}},0}}}
\newcommand{\mhgstatnull}{{s^{\textrm{\tiny{mHG}},0}}}

\newcommand{\mhgp}{p^{\textrm{\tiny{mHG}}}}

\newcommand{\xlmhgstat}{{s^{\textrm{\tiny{mHG}}}_{\textsc{x,l}}}}
\newcommand{\xlmhgstatnullrnd}{{S^{\textrm{\tiny{mHG}},0}_{\textsc{x,l}}}}

\newcommand{\xlmhgp}{{\mhgp_{\textsc{x,l}}}}

\newcommand{\VKN}{{\mathcal{V}^{(K,N)}}}
\newcommand{\VKNex}{{\mathcal{V}^{(20,100)}}}

\newcommand{\MKN}{{\mathcal{M}^{(K,N)}}}
\newcommand{\MKNex}{{\mathcal{M}^{(20,100)}}}

\newcommand{\RXL}{{\mathcal{R}_{\textsc{x,l}}}}
\newcommand{\Rmod}{{\mathcal{R}'}}

\newcommand{\knnull}{{k_{(n)}^0}}
\newcommand{\emax}{{e^{\max}}}
\newcommand{\esimple}{{e_{(n)}}}
\newcommand{\nstar}{{n^{*}}}
\newcommand{\estar}{{e^{*}}}
\newcommand{\enstar}{{e_{(n^{*})}}}

\newcommand{\mhge}{{e^{\textrm{\tiny{mHG}}}}}
\newcommand{\xlmhge}{{e^{\textrm{\tiny{XL-mHG}}}_{\textsc{x,l}}}}

\newcommand{\Cpsi}{{\mathcal{C}(\psi)}}
\newcommand{\mhgepsi}{{e^{\textrm{\tiny{mHG}}}\,(\psi)}}
\newcommand{\Cxlpsi}{{\mathcal{C}_{\textsc{x,l}}(\psi)}}
\newcommand{\xlmhgepsi}{{e^{\textrm{\tiny{mHG}}}_{\textsc{x,l}}\,(\psi)}}

\DeclareMathOperator*{\argmin}{arg\,min}

\title{The XL-mHG Test For Enrichment:\\A Technical Report}
\author[\empty]{Florian Wagner\textsuperscript{1,}\thanks{Email: \url{florian.wagner@duke.edu}}}
\affil{\small{\textsuperscript{1}PhD Program in Computational Biology and Bioinformatics,\newline Duke University}}
\date{}

\begin{document}

\maketitle

\begin{abstract}
The minimum hypergeometric test (mHG) is a powerful nonparametric hypothesis test to detect enrichment in ranked binary lists. Here, I provide a detailed review of its definition, as well as the algorithms used in its implementation, which enable the efficient computation of an exact p-value. I then introduce a generalization of the mHG, termed XL-mHG, which provides additional control over the type of enrichment tested, and describe the precise algorithmic modifications necessary to compute its test statistic and p-value. The XL-mHG algorithm is a building block of GO-PCA, a recently proposed method for the exploratory analysis of gene expression data using prior knowledge.
\end{abstract}

\pagebreak
\tableofcontents

\pagebreak

\section{Introduction}\label{sec:intro}

On the face of it, the \textit{minimum hypergeometric} test, or simply \textit{mHG}, is a nonparametric hypothesis test to detect enrichment in ranked binary lists \cite{eden_discovering_2007}. This may sound more exotic than it actually is, since what the mHG essentially provides is a powerful way of testing for a (directed) association between one continuous and one binary variable, while making very few assumptions about the distributional properties of either variable (for a more detailed discussion of the mHG as a test for association, see \cref{sec:mhg_assoc}). The mHG is therefore \textit{very} generally applicable. So far, however, the mHG has mostly been applied to biological problems, e.g. the detection of DNA sequence motifs in transcription factor binding sites \cite{eden_discovering_2007}, or the detection of enriched Gene Ontology (GO) terms in ranked lists of genes \cite{eden_gorilla:_2009}.

The XL-mHG is an extension of the mHG that introduces two parameters ($X$ and $L$), which are designed to provide additional control over the minimal subset size that can constitute enrichment ($X$), and the part of the list that is to be tested for enrichment ($L$). Depending on the application, these parameters can help to significantly increase the specificity of the test. This report provides a detailed description of how to efficiently implement the mHG, and its extension, the XL-mHG, thus allowing for a single test to be performed in milliseconds, even for lists containing thousands of elements.

This manuscript is organized as follows: \cref{sec:mhg} provides a review of the mHG for readers that are not familiar with it, assuming no background knowledge other than a familiarity with basic concepts of probability theory. \cref{sec:mhgimpl} provides a detailed discussion of how to efficiently implement the mHG. This discussion is based entirely on ideas developed by Dr. Zohar Yakhini and colleagues \cite{eden_discovering_2007}. \cref{sec:xlmhg} introduces an extension of the mHG, termed XL-mHG, which was designed to provide additional control over the type of enrichment that is tested for. This extension was developed by me, and is used in a recently proposed biological application for knowledge-based unsupervised analysis of heterogeneous expression data \cite{wagner_go-pca:_2015}. \cref{sec:xlmhg} also includes a detailed discussion of how to modify the efficient mHG algorithms to calculate XL-mHG test statistics and their associated p-values. Certain algorithms and derivations are provided in the \hyperref[sec:appendix]{Appendix}. A free and open-source \href{http://cython.org/}{Cython} implementation of the XL-mHG can be found at \url{https://github.com/flo-compbio/xlmhg}.

\pagebreak

\section{The minimum hypergeometric (mHG) test}\label{sec:mhg}

The mHG is an enrichment test for ranked binary lists that was developed by Dr. Zohar Yakhini and colleagues \cite{eden_discovering_2007}. This section serves as a review of the mHG for readers that are not familiar with it. I first introduce the representation of ranked binary lists as binary vectors in \cref{sec:list_rep}. Then, in \cref{sec:simple_enrich}, I describe a simpler enrichment test for such lists, and demonstrate its application on a toy example. The discussion of this simple test serves multiple purposes: First, the simple test is directly related to the mHG through its reliance on the \hyperref[sec:hypergeom]{hypergeometric distribution}, and almost all of the notation and concepts introduced in this section serve as the basis for my later discussion of the mHG. Second, I highlight the fact that the simple test suffers from a major drawback, which the mHG was specifically designed to overcome. The discussion of the simple test therefore provides a strong motivation for the following discussion of the mHG. A discussion of how to efficiently implement the mHG is postponed to \cref{sec:mhgimpl}.

\subsection{Representing a ranked binary list as a vector}\label{sec:list_rep}

We represent the ranked binary list we are talking about as a vector $\bm{v}$ of length $N$, with entries of only zeros and ones:

\begin{align*}
\bm{v} = (v_1,v_2, \dots, v_N)^T, \: v_i \in \{0,1\}
\end{align*}

We refer to individual entries in this list as ``elements'' (adopting the terminology used for vectors). We refer to the set of all elements for which $v_i = 0$ as ``the 0's'', and to the set of all other elements as ``the 1's''. We also say that $v_1$ represents the ``topmost'' element, and $v_N$ the ``bottommost'' element of the list. We further let $K$ and $W$ denote the total number of 1's and 0's in the list, respectively ($K + W = N$):

\begin{align*}
K &= \sum_{i=1}^N \ind[v_i = 1]  \\
W &= \sum_{i=1}^N \ind[v_i = 0] = N - K
\end{align*}

Here, $\ind[\,]$ denotes the \textit{indicator function}. We can think of the 1's as representing the elements with an ``interesting'' feature in the list. For example, if we are dealing with a list of genes, the genes represented by 1's might represent all genes that are known to play a role in DNA replication. Typically, the number of 1's is smaller (and sometimes much smaller) than the number of 0's ($K < W$ or $K \ll W$).

\subsection{A simple test for enrichment (with a major drawback)}\label{sec:simple_enrich}

For demonstration purposes, we will assume that we are given a particular vector $\vex$, representing a ranked binary list (as explained above). The vector looks as follows:
\begin{align*}
\vex = (1,0,1,1,0,1,0,0,0,0,0,0,0,0,0,0,0,0,1,0)^T
\end{align*}
Using the notation introduced above, we have $N=20$, $K=5$, and $W = 20-5 = 15$. We are interested in whether there is an enrichment of 1's ``at the top of the list''. Of course, in order to be able to provide a quantitative answer to this question, we need to first define what exactly we mean when we say ``enrichment at the top of the list''.

One possibility is to directly define ``the top of the list'' by introducing an integer ``cutoff'' parameter $n$ ($0$\textless{}$n$\textless{}$N$), indicating that ``the top of the list'' consists of the first $n$ elements of $\bm{v}$. To quantify enrichment, we would then calculate a \textbf{test statistic} $\kn$, representing the number of 1's we observe among the first $n$ elements:

\begin{align*}
\kn = \sum_{i=1}^n \mathds{1}[v_i = 1]
\end{align*}
Under the null hypothesis of no enrichment, we assume that the 0's and 1's are randomly distributed in the list (in other words, we assume that all permutations of $\bm{v}$ are equally likely). We can then use the \textit{hypergeometric distribution} to assess the \textbf{statistical significance} of observing a certain $\kn$.

\vspace{1em}
\fbox{\begin{minipage}{0.9\textwidth}

\subsection*{The hypergeometric distribution}\label{sec:hypergeom}

A random variable has the hypergeometric distribution if it represents the number $k$ of ``interesting'' items (here, the ``1's'' in the list) among a sample of size $n$, drawn without replacement from a population of $N$ items, of which $K$ are considered ``interesting''. The probability mass function (PMF) of the hypergeometric distribution is defined as follows:

\begin{equation}\label{eq:hypergeom}
\tag{Hypergeometric PMF}
\fk = \frac{\binom{K}{k}\binom{N-K}{n-k}}{\binom{N}{n}}
\end{equation}

We use $f$ to represent the hyergeometric PMF throughout this document. In the PMF, $\binom{a}{b}$ denotes the \textit{binomial coefficient} (read as ``a choose b''). It represents the number of ways in which we can select $b$ elements from a total of $a$ elements (${b \leq a}$), when we ignore the order in which we choose the elements. The binomial coefficient can be calculated as follows:

\begin{align*}
\binom{a}{b} = \frac{a!}{b!(a-b)!}
\end{align*}
\end{minipage}}
\vspace{1em}

To assess whether $\kn$ is statistically significant, we need to calculate its associated \textbf{p-value} $\hgp$. This p-value is defined as the probability of observing \underline{$\kn$ or more 1's} among the first $n$ items in the list, under the assumption that the 1's and 0's are randomly distributed. This probability can be calculated as the ``tail'' of the hypergeometric distribution:

\begin{align}
\hgp{} &= \sum_{i=\kn}^{\min(n,K)} f(i;\, N,K,n) \label{eq:hgpsum} \\
                &= 1 - F(\kn-1;\, N,K,n) \nonumber \\
                &= S(\kn-1;\, N,K,n) \nonumber
\end{align}
Here, $F$ and $S$ denote the cumulative density function (CDF) and the survival function (SF) of the hypergeometric distribution, respectively:

\begin{align*}
F(k; N,K,n) &= \sum_{i=0}^{k} f(i; N,K,n)\\
S(k; N,K,n) &= \sum_{i=k+1}^{\min(n,K)} f(i; N,K,n)
\end{align*}

For almost every commonly used programming framework, there are publicly available packages that provide functions for evaluating $f$, $F$, and/or $S$. For example, in Python, the \texttt{SciPy} package offers the functions \texttt{stats.hypergeom.sf} for evaluating $S$. It is therefore straightforward to implement this test in most circumstances. Given a specific $n$, we can now calculate a \textbf{p-value} $\hgp$ for our test statistic $\kn$. Since the length of $\vex$ is $N=20$, we might choose $n=N/4=5$, corresponding to the top 25\% of elements, as a reasonable choice for the ``top of the list''. We then have:

\begin{align*}
\kex{} &= \sum_{i=1}^5 \mathds{1}[v_i = 1] = \uuline{3}\\
\hgpex{} &= S(3-1,20,5,5) \approx \uuline{0.073}
\end{align*}
The result of this calculation shows that at the conventional significance level of $\alpha = 0.05$, we \textit{can't} reject the null hypothesis of no enrichment (since $\hgpex > \alpha$). This result might seem counter-intuitive: Doesn't the distribution of 1's in our vector look highly skewed towards the top (except for one outlier at the bottom)? Here is $\vex$ again, with the first 5 elements visually separated to indicate the ``top of the list'' for our cutoff of $n=5$:

\begin{align*}
\vex = (1,0,1,1,0, |\; 1,0,0,0,0,0,0,0,0,0,0,0,0,1,0)^T
\end{align*}

The reader may have noticed that the element at our chosen cutoff of $n = 5$ is a 0 located right ``in between'' two 1's. What if we had instead chosen $n=4$, excluding the 0, or $n=6$, including another 1? We would then have:

\begin{align*}
\hgpexa &\approx \uuline{0.032}\\
\hgpexb &\approx \uuline{0.014}
\end{align*}
What these results show is that, had we chosen either $n=4$ or $n=6$, we actually \textit{would} have been able to reject our null hypothesis of no enrichment! This example illustrates a \textbf{major drawback} of our simple enrichment test: \textit{The choice of $n$ strongly affects the outcome of the test}. This is a big problem in practice, because we often do not have a way to determine the ``best'' $n$ to use. If we simply choose $n$ arbitrarily (e.g., $n=N/4$, as in our example), then there will be situations where we choose $n$ too small, meaning that we miss a surprising accumulation of 1's that occurs relatively high in the list, but below the n'th rank. But we also do not want to choose $n$ too large, since that could result in an insignificant p-value when there actually are a surprising number of 1's concentrated at the very top of the list.

What about other ideas for choosing $n$? Why not try to first look at $\bm{v}$, and choose an $n$ that appears to ``capture'' a large number of 1's? In our toy example, this approach seemed to ``work wonders'': A blind choice of $n=5$ did not result in a significant test, but a ``data-driven'' choice of $n=4$ or $n=6$ did! While this idea might seem attractive at first, it is actually very problematic, since the choice of $n$ using this method is subjective. Suppose two people ``eyeball'' different $n$'s, resulting in one significant and one insignificant test. Then, the question of whether there is or isn't statistically significant enrichment would fundamentally come down to a question over who has the better judgement. Clearly, this would not be a scientifically sound method. A statistical way of describing essentially the same problem is to say that choosing $n$ after taking a ``peek'' at the data introduces a significant amount of (positive) \textit{bias} to the test, so that the p-values obtained would overstate the statistical significance of the observed pattern. Moreover, since the choice is subjective, it would be impossible to quantify the extent of this bias, making the test all but statistically useless.

Finally, we could decide to try a few different $n$'s (without looking at the data), and see if any of those tests come back significant. However, this would then constitute \textit{multiple testing}, which we would then have to account for (raising new problems). In short, none of these ideas provide a useful approach for the criticial choice of $n$ when it is not known \textit{a priori}.

\subsection{mHG: A nonparametric test for enrichment}\label{sec:mhg_nonparam}

In \cref{sec:simple_enrich}, we discussed an enrichment test for ranked binary lists that is simple to implement, but suffers from the major drawback of requiring upfront knowledge of $n$, the parameter that defines what part of the list should be considered ``the top''. The mHG provides a very elegant solution to this problem, by giving up on the idea of defining ``the top of the list'' altogether. The mHG gets rid of the $n$ parameter, and replaces it with --- nothing. The mHG does not require \textit{any} parameters; in this sense, it is a fully \textit{nonparametric} test. In cases where $n$ is unknown (i.e., most of the time), this property provides the mHG with a \textit{huge} advantage over the simple enrichment test described above. However, to achieve this advantage, the mHG relies on a more complex testing procedure, which requires many ``sub-tests'' to be performed. This in turn introduces a \textit{multiple testing} problem\footnote{For a single hypothesis test, its associated p-value represents the probability of rejecting the null hypothesis when it is in fact true. The multiple testing problem refers to the fact that when many hypothesis tests are performed simultaneously, their individual p-values no longer represent that probability ---  with enough ``tries'', we always expect to obtain some ``significant'' p-values, even if all hypotheses tested are truly null. This means that the testing procedure must somehow be \textit{corrected} for the fact that many individual tests were performed.}. Fortunately, it turns out that these issues can be resolved very efficiently.

\begin{figure}[!ht]
        \centering
        \includegraphics[width=0.3\textwidth]{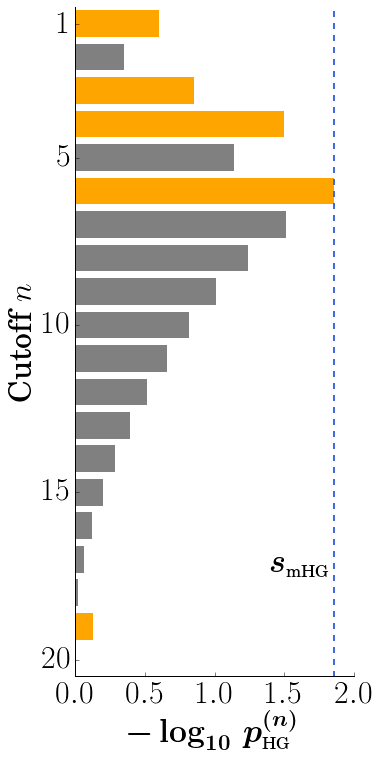}
        \caption{Calculation of the mHG test statistic $\mhgstat$ for the ranked binary list $\vex$ (the example used in the text). Each bar represents the hypergeometric p-value $\hgp$ at a given cutoff $n$. Bars highlighted in yellow correspond to positions in the list that have a ``1'' element. $\mhgstat$ corresponds to the smallest $\hgp$ (represented here by the largest bar, due to the negative log scale). In this example, $\mhgstat$ occurs at a cutoff of $n=6$ (where $\kn = 4$). Note that for each cutoff where the list contains a ``0'', $\hgp$ is always larger than $\hgpm$ (represented by a smaller bar). We can therefore skip the calculation of $\hgp$ for those $n$ when we calculate $\mhgstat$.}
        \label{fig:mhg_stat}
\end{figure}

The mHG method consists of two components: The first component is the definition of the \textbf{mHG test statistic} $\mhgstat$. For the simple enrichment test described above, the test statistic was simply $\kn$, the number of 1's among the first n elements of the list. However, without a fixed cutoff $n$, this statistic obviously does not apply. Instead, $\mhgstat$ is defined as the minimum hypergeometric p-value $\hgp$, taken over \textit{all possible cutoffs} (see \cref{fig:mhg_stat}):
\begin{align}\label{eq:mhgstat}
\tag{mHG test statistic}
\mhgstat = \min_n \hgp
\end{align}
Note that due to this definition, \textit{smaller} values of $\mhgstat$ represent \textit{stronger} enrichment. 


The second component of the mHG is a way of efficiently calculating an \textbf{exact p-value} $\mhgp$ for $\mhgstat$. The null model is still the same as before: We assume that when there is no enrichment, all permutations of $\bm{v}$ are equally likely. However, we have no closed form solution for the distribution of $\mhgstat$ under this model. Moreover, there exist an astronomically large number of permutations, even for moderately-sized lists (say, N=100). This means that we cannot calculate a p-value by simply enumerating all those permutations and calculating their test statistic. Therefore, the exact calculation of $\mhgp$ relies on a dynamic programming approach for \textit{path counting} \cite{eden_discovering_2007} (discussed in \cref{sec:mhg_pvalue}).

In addition to describing how to calculate $\mhgp$ exactly and in polynomial time, Eden et al. also derived a useful upper bound for $\mhgp$, known as the \textit{Lipson bound} \cite{eden_discovering_2007}. Its form is surprisingly simple:
\begin{align}\label{eq:mhgbound}
\tag{Lipson bound}
    \mhgp \le K \mhgstat
\end{align}

The argument used to derive this bound relies on the fact that even though $\mhgstat$ is defined as the minimization over the p-values of $n$ hypergeometric tests, the mHG test can be shown to be equivalent to testing no more than $K$ distinct ``key'' cutoffs $n_k$, one for each $k \in \{1\dots{}K\}$. We typically do not know or care about the exact values of these $n_k$, but we can still use this fact in order to derive the bound.

\subsection{The notion of enrichment behind the mHG test}

What notion of enrichment does the \hyperref[eq:mhgstat]{definition} of the mHG test statistic translate to? At first glance, this might seem like an obvious question, since the mHG test statistic is indeed just the minimization of the simple hypergeometric test's p-value $\hgp$ over all possible cutoffs $n$, and the notion of enrichment underlying the simple hypergeometric test is easily described: The test asks, ``Is there a surprisingly large number of 1's above the cutoff $n$?'' A key property of the test is that it does not take into account the exact distribution of 1's and 0's above and below the cutoff, only their counts. For example, some 1's could be located at the very bottom of the list, yet the significance of the test would be just the same if those 1's were located right below the cutoff. This makes the test robust to (negative) outliers, and is in contrast to other nonparametric tests, such as the Mann-Whitney U test, which operate on the precise ranks of the 1's and 0's. However, even in less extreme cases, it is important to understand that the behavior of a subset of 1's (i.e. their location in the list) can lead to a positive test result, while the behavior of the remaining 1's is essentially ignored. In particular, when $n$ is small in relation to the length of the list, the subset of 1's responsible for a positive test result can represent only a small fraction of the total number of 1's. In other words, for relatively small $n$, a positive test result can be based on a small fraction of ``interesting'' items that are located at the very top of the list. In other situations, when $n$ is large, a positive test can have the opposite interpretation, as it can be based on a slight overrepresentation of 1's above the cutoff $n$.

The point of this discussion is to emphasize that the type of enrichment detected by the simple hypergeometric test can vary significantly, depending on the choice of the cutoff parameter $n$. However, every single mHG test considers all possible $n$ simultaneously. What does this mean for the notion of enrichment underlying the mHG? One way of addressing this question from an intuitive standpoint is to imagine that the mHG inherits \textit{all} aforementioned (and contradictory) behaviors of the simple hypergeometric test, while the choice of which of these behaviors is ``expressed'' in any specific case depends on the tested list itself:\footnote{The reader may excuse my use of a biological term in this context.} For lists where a small subset of 1's is located at the very top, the mHG will likely base its enrichment on this subset; for lists with only a slight overrepresentation of 1's among, say, the first half of the list, the mHG will detect enrichment based on that pattern. One could therefore say that the notion of enrichment behind the mHG is ``data-dependent''. This of course is precisely why the mHG is so useful: It ``adapts'' the general notion of enrichment to the specific situation encountered. In \cref{sec:xlmhg}, I introduce two parameters which, abstractly speaking, provide additional control over \textit{how much} the mHG is allowed to adapt its notion of enrichment.

\subsection{Testing for directed association}\label{sec:mhg_assoc}

Before moving on to a description of how to efficiently implement the mHG, I would like to emphasize that besides testing for \textit{enrichment}, the purpose of the mHG can be more generally understood as testing for \textit{association} between one continuous and one binary feature: The continuous feature is used to rank all items, and the binary feature marks the ``interesting'' items (the 1's), as discussed above. Moreover, this test is \textit{directional}: The mHG only tests for enrichment at the top of the list, not at the bottom. Therefore, an accumulation of 1's at the bottom of the list does not result in a significant test. Of course, enrichment at the bottom can be tested separately, by simply inverting the list.

\pagebreak

\section{Efficient implementation of the mHG test}\label{sec:mhgimpl}

In this section, I provide a detailed description of how to efficiently calculate the mHG test statistic and its associated p-value, for any ranked binary list. The ideas for this implementation were developed by Dr. Zohar Yakhini and colleagues \cite{eden_discovering_2007}. I present them here in full detail, which then allows me (in \cref{sec:xlmhg}) to precisely describe the modifications required to accomodate the two parameters introduced by the XL-mHG test statistic.

\subsection{Calculating the mHG test statistic $\mhgstat$}\label{sec:mhgimpl_stat}

The \ref{eq:mhgstat} $\mhgstat$ is defined as the minimum hypergeometric p-value $\hgp$, taken over all possible cutoffs. In principle, the calculation of $\mhgstat$ is therefore very simple:
\begin{algorithm}[H]
\begin{algorithmic}
\STATE $\mhgstat \leftarrow 1.0$
\FOR{all $n$}
\STATE Calculate $\hgp$
\STATE $\mhgstat \leftarrow \min \{\hgp,\mhgstat\}$
\ENDFOR
\RETURN $\mhgstat$
\end{algorithmic}
\end{algorithm}
However, for large $N$, calculating the values of all the $\hgp$ individually is relatively slow. To calculate $\mhgstat$ more efficiently, we can rely on two key observations: First, we know that the smallest $\hgp$ will never occur at a cutoff $n$ for which $v_n = 0$. We can therefore skip the calculation of $\hgp$ for all ``0'' elements, which leads to a significant speed-up when ${K \ll N}$. (For similar reaons, we can also skip the calculation of $\hgp$ when $v_{n+1} = 1$, which leads to significant speed-up when there are long stretches of consecutive 1's in the list.) Second, even when we hit a 1, we can avoid calculating $\hgp$ ``from scratch''. Instead, we can exploit the fact that $N$ and $K$ remain constant throughout the procedure, and use a recursive approach to efficiently calculate all the $\hgp$ for which ${v_n = 1}$. This approach consists of two sub-algorithms: \cref{alg:hgp_from_fkn} calculates $\hgp$ from $\fkn$ in $\mathcal{O}(K)$, and \cref{alg:fkn_recursive} calculates $\fkn$ for all $n$ in $\mathcal{O}(N)$. \cref{alg:mhgstat} combines these sub-routines to calculate $\mhgstat$ in $\mathcal{O}(KN)$.

\subsection{Calculating the mHG p-value $\mhgp$}\label{sec:mhg_pvalue}

Suppose that we have calculated the value of $\mhgstat$ for a ranked binary list $\bm{v}$ with $N$ elements and $K$ 1's. How is its associated p-value defined? Recall that under the null model, all permutations of $\bm{v}$ are assumed to be equally likely. Therefore, let $\mhgstatnullrnd$ be a random variable that represents the mHG test statistic obtained for a random permutation of $\bm{v}$. The p-value $\mhgp$ associated with $\mhgstat$ is then defined as the probability of observing an $\mhgstatnullrnd$ that is at least as good (i.e., smaller than or equal to, see \cref{sec:mhgimpl_stat}) as $\mhgstat$:
\begin{align}\label{eq:mhg_pval}
\tag{mHG p-value}
\mhgp = \Pr(\mhgstatnullrnd \le \mhgstat)
\end{align}

Let $\VKN$ represent the set of all possible permutations $\vnull$ of $\bm{v}$ (including $\bm{v}$ itself), i.e., \textit{all} ranked binary lists with $N$ elements and exactly $K$ 1's. Then, let $\vnullrnd$ be a random variably representing a list randomly drawn from $\VKN$. Using the \textit{fundamental bridge}\footnote{A term used by Harvard Professor Joe Blitzstein for the relationship between probabilities and expectations of indicator random variables, see \url{http://www.quora.com/What-are-the-top-10-big-ideas-in-Statistics-110-Introduction-to-Probability-at-Harvard}}, we can then re-express $\mhgp$ as follows:
\begin{align*}
    \mhgp   &= \mathbb{E}\,\big(\,\ind[\mhgstatnull \le \mhgstat\: |\; \vnullrnd]\,\big) \\
            &= \bigg(\,\sum_{\vnull \in \VKN} \mathds{1}[\mhgstatnull \le \mhgstat]\,\bigg) \: /\; |\,\VKN\,|
\end{align*}
This would then suggest the following (naive) algorithm:
\begin{algorithm}[H]
\begin{algorithmic}
\STATE $p \leftarrow 0$
\FOR{each $\,\vnull \in \VKN$}
\STATE Calculate $\mhgstatnull$ (the mHG test statistic for $\vnull$)
\IF{$\mhgstatnull \le \mhgstat$}
\STATE $p \leftarrow p+1$
\ENDIF
\ENDFOR
\RETURN $\mhgp = p\, /\, |\VKN|$
\end{algorithmic}
\end{algorithm}
Unfortunately, the number of lists in $\VKN$ grows incredibly quickly:
\begin{align*}
|\VKN| = \binom{N}{K} = \frac{N!}{K!(N-K)!}
\end{align*}
For example, $|\VKNex| \approx 5.4 \times 10^{20}$. In this case, we would therefore have to calculate more than $10^{20}$ (!) different $\mhgstatnull$ in order to calculate $\mhgp$, which shows that this approach is completely infeasible, except for \textit{very} short lists.

Instead, the efficient calculation of $\mhgp$ relies on the idea of \textit{path counting} \cite{eden_discovering_2007}. To understand this idea, let us first take a step back and again look at the definition of $\mhgstat$:
\begin{align*}
\mhgstat &= \min_n \hgp
\end{align*}
We also saw (in \cref{sec:simple_enrich}) that each $\hgp$, in turn, can be calculated as:
\begin{align*}
\hgp &= S(\kn-1;\, N,K,n)
\end{align*}
This shows that the value of each $\hgp$ depends on exactly four parameters: n, $\kn$, $N$, and $K$. Note that all $\bm{v}^0$ share the same $N$ and $K$. Therefore, the only parameters that vary during the calculations of their $\mhgstatnull$ are $n$, the cutoff, and $\kn$, the number of 1's above the cutoff. How many unique parameter combinations $(\kn,n)$ are there? We know that $0 \le \kn \le n$ and $0 \le n \le N$. Therefore, there are less than ${(K+1)}\times{}{(N+1)}$ unique combinations. This leads us to a surprising observation: Despite the fact that there are more than $10^{20}$ unique $\bm{v}^0$ in $\VKNex$, the calculations of \textit{all} of their mHG test statistics $\mhgstatnull$ depend on less than $21\times101=2121$ unique values for $\hgp$!


\begin{figure}[!ht]
        \centering
        \includegraphics[width=0.9\textwidth]{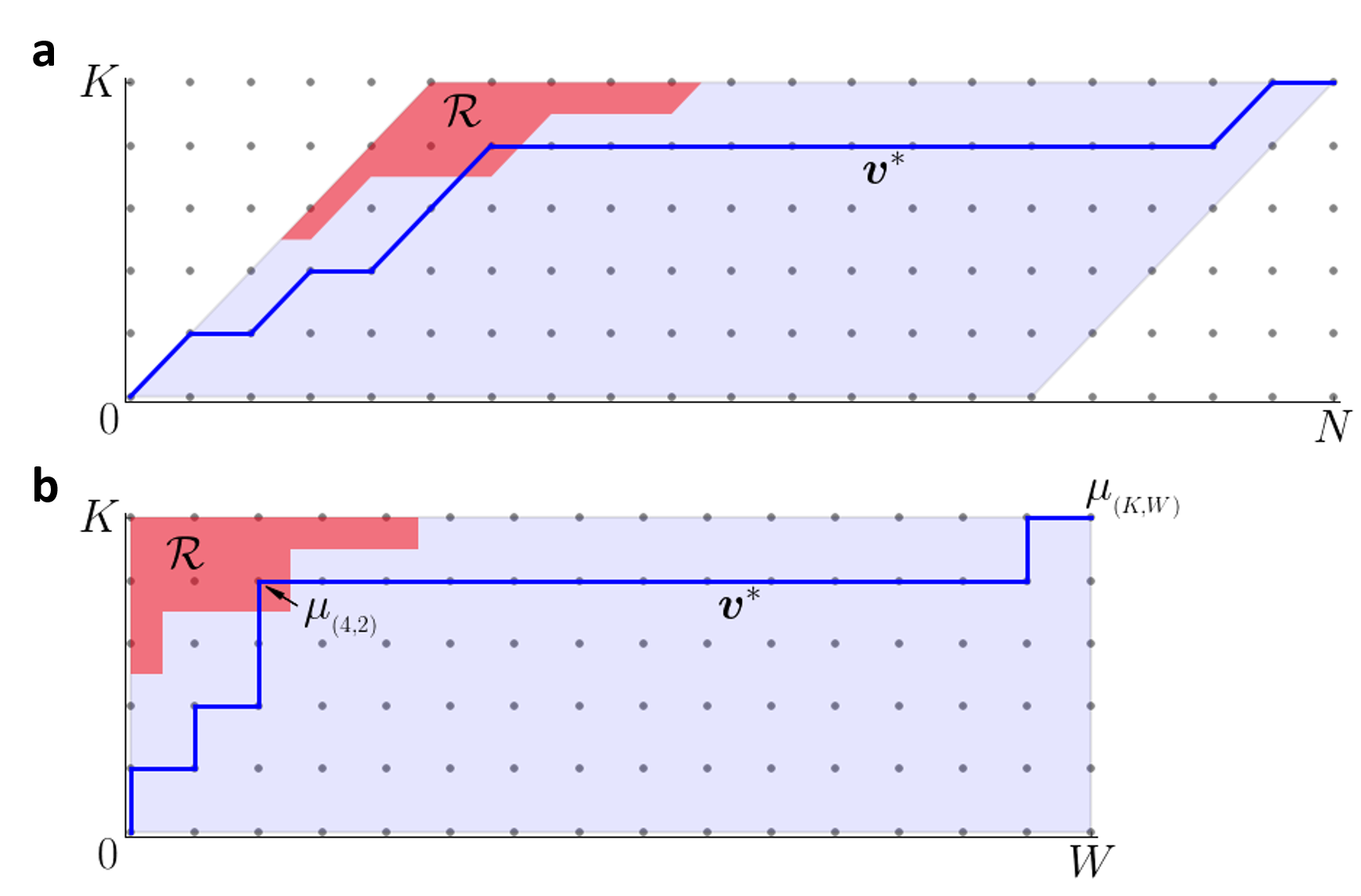}
        \caption{The set $\MKN$ of all hypergeometric configurations. \textbf{a} A ${(K+1)}\times{(N+1)}$ grid showing $\MKN$ as the blue shaded area. Each $\vnull \in \mathcal{V}^{N,K}$ can be represented as a unique \textit{path} through $\MKN$. The path for the example vector $\bm{v}^*$ is shown in blue. The set $\mathcal{R}$ of all configurations with an mHG test statistic as good or better than that of $\bm{v}^*$ is shown in red. Note that the points within the white areas do not represent valid configurations (cf. Figure 7 in \cite{eden_discovering_2007}). \textbf{b} A more compact ${(K+1)}\times{(W+1)}$ grid for representing $\MKN$. There exists a 1-to-1 mapping between the configurations $\mu'_{(k,n)}$ in \textbf{a} and the configurations $\mu_{(k,w)}$ in \textbf{b}.}
        \label{fig:mhg_pval}
\end{figure}

Since a parameter combination $(k,n)$ uniquely determines the value of the hypergeometric p-value (assuming constant $K$ and $N$), we refer to it as a \textit{hypergeometric configuration} $\mu'_{(k,n)}$. We can then define the set of all hypergeometric configurations $\MKN = \{\mu'_{(k,n)}: 0 \le k \le n,\, 0 \le n \le N\}$. We can visualize this set using a ${(K+1)}\times{(N+1)}$ grid (see \cref{fig:mhg_pval}a). However, we can simplify our indexing with $W = N-K$ and $w = n-k$. ($W$ represents the total number of 0's in the list, and $w$ the number of 0's above the cutoff $n$.) We can then use $k$ and $w$ to define hypergeometric configurations $\mu_{(k,w)}$, allowing us to equivalently define $\MKN$ as follows:
\begin{align*}
\MKN = \{\mu_{(k,w)}: 0 \le k \le K,\, 0 \le w \le W\}
\end{align*}

The fact that both definitions of $\MKN$ describe the same set of hypergeometric configurations should be obvious by comparing \cref{fig:mhg_pval}a to \cref{fig:mhg_pval}b. We can now easily calculate $|\MKNex| = (20+1)\times((100-20)+1) = 1701$. Therefore, there are exactly 1701 unique hypergeometric configurations involved in the calculations of the $\mhgstatnull$ for all the $\vnull$ in $\VKNex$. For each configuration $\mu_{k,w}$, let $\pkw$ represent its associated hypergeometric p-value:
\begin{align*}
\pkw = S(k-1;\, N,K, k+w)
\end{align*}
We can then define the set $\mathcal{R}$ of all configurations with a hypergeometric p-value at least as good as $\mhgstat$:
\begin{align*}
\mathcal{R} = \{\mu_{(k,w)} : p_{(k,w)} \le \mhgstat\}
\end{align*}
Importantly, we can determine whether $\mu_{(k,w)} \in \mathcal{R}$, for all $\mu_{(k,w)}$, in $\mathcal{O}(KW)$, using \cref{alg:mhg_pval_R}.

We make another observation relating $\VKN$ to $\MKN$: Each ${\vnull \in \VKN}$ has a unique representation as a \textit{path} ${\bm{\lambda}^0 = (\mu_0,\mu_1,\dots,\mu_N)}$, consisting of all the hypergeometric configurations $\mu_n$ that we encounter when we go over all cutoffs $n$ for $\vnull$. Using our ${(k,w)}$-indexing scheme, all paths start with $\mu_0 = \mu_{(0,0)}$ and end with $\mu_N = \mu_{(k,w)}$. \cref{fig:mhg_pval} shows the path representation of $\vex$. We say that a path ``crosses $\mathcal{R}$'' when at least one of its $\mu_n$ is in $\mathcal{R}$. We can then express $\mhgp$ in terms of path counts:
\begin{align*}
\mhgp   &= \frac{\textrm{\# of paths that cross $\mathcal{R}$}}{\textrm{total \# of paths}} \\
        &= 1 - \frac{\textrm{\# of paths that \textit{don't} cross $\mathcal{R}$}}{\textrm{total \# of paths}}
\end{align*}


\begin{figure}[!ht]
        \centering
        \includegraphics[width=0.9\textwidth]{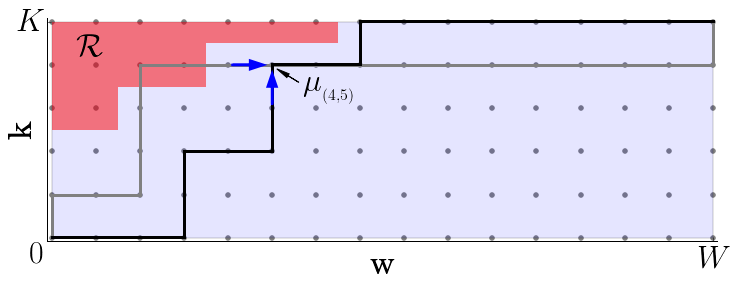}
        \caption{Counting the number of paths that \textit{don't} cross $\mathcal{R}$ using a dynamic programming approach \cite{eden_discovering_2007}. All hypergeometric configurations $\mukw$ are represented on an $(K+1)\times(W+1)$ grid, as in \cref{fig:mhg_pval}. Two paths are shown, one of which crosses $\mathcal{R}$ (shown in gray). All paths containing $\mu_{(4,5)}$ also contain either $\mu_{(4,4)}$ or $\mu_{(3,5)}$ (blue arrows).}
        \label{fig:mhg_pval_recur}
\end{figure}

Using these observations and definitions, it is possible to calculate $\mhgp$ without explicitly calculating $\mhgstatnull$ for each $\vnull \in \VKN$. Instead, we can rely on a \textit{dynamic programming} approach to count the number of paths that don't cross $\mathcal{R}$. This approach relies on the observation that all paths that contain a certain configuration $\mukw$ (k $>$ 0, w $>$ 0) also contain either $\mukwmk$ or $\mukwmw$ (see \cref{fig:mhg_pval_recur}). Let $\ckw$ represent the fraction of paths that contain $\mukw$:
\begin{align*}
\ckw = |\{\bm{\lambda}^0: \mukw \in \bm{\lambda}^0\}|\, /\, |\VKN|
\end{align*}
We then observe the following recurrence relation for $\ckw$:
\begin{align*}
\ckw = \ckwmk \frac{K-k+1}{N-n+1} + \ckwmw \frac{W-w+1}{N-n+1}
\end{align*}

Using this recurrence relation, it is straightforward to recursively calculate the fraction of paths that don't cross $\mathcal{R}$. We let $\mkw$ represent the fraction of paths that don't cross $\mathcal{R}$, but contain the configuration $\mukw$:
\begin{align*}
\mkw = |\{\bm{\lambda}^0: \mukw \in \bm{\lambda}^0 \textrm{, and } \mu \notin \mathcal{R} \textrm{ for all } \mu \in \bm{\lambda}^0\}|\, /\, |\VKN|
\end{align*}
We have:
\begin{equation*}
\mkwstart =
\begin{cases}
0, & \textrm{if } \mhgstat = 1.0 \\
1.0 & \textrm{otherwise}
\end{cases}
\end{equation*}
Then we observe the following recurrence relation for the $\mkw$ (k$>$0, w$>$0):
\begin{equation*}
\mkw =
\begin{cases}
0, & \textrm{if }\mukw \in \mathcal{R} \\
\mkwmk \frac{K-k+1}{N-n+1} + \mkwmw \frac{W-w+1}{N-n+1} & \textrm{otherwise}
\end{cases}
\end{equation*}
If k=0, or w=0, the first or second term of the recurrence relation is omitted, respectively, for the case ${\mukw \notin \mathcal{R}}$. \cref{alg:mhg_pval} uses this relation to calculate $m_{(K,W)}$ in $\mathcal{O}(KW)$, yielding $\mhgp$:
\begin{align*}
\mhgp = 1 - m_{(K,W)}
\end{align*}

\pagebreak

\section{The extended mHG (XL-mHG) test}\label{sec:xlmhg}

In this section, I will first discuss two limitations of the mHG (in \cref{sec:xlmhg_limit}), in order to then motivate the definition of the XL-mHG test statistic, which involves two new parameters (see \cref{sec:xlmhg_stat}), as well as the XL-mHG p-value (see \cref{sec:xlmhg_pval}). In both cases, I will include a discussion on how to modify the efficient algorithms used in the implementation of the mHG (see \cref{sec:mhgimpl}), in order to accommodate the two new parameters.

\subsection{Limitations of the mHG test}\label{sec:xlmhg_limit}

As previously discussed (see \cref{sec:mhg_nonparam}), abandoning the cutoff parameter $n$ provides the mHG with an enormous advantage over simpler tests. However, the approach used by the mHG represents the ``other extreme'', in the sense that the mHG does not exert \textit{any} control over which cutoffs are tested for enrichment. In certain scenarios, this lack of control can turn into an ``Achilles heel'' and significantly reduce the usefulness of the test.

\begin{figure}[!ht]
        \centering
        \includegraphics[width=0.9\textwidth]{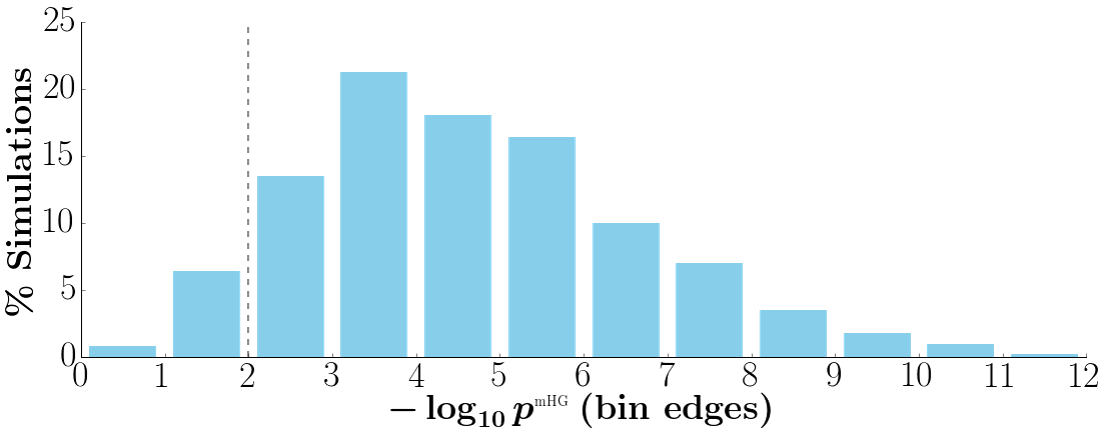}
        \caption{Assessing the sensitivity of the mHG to weak enrichment within the top 50\% of a long list (Scenario 1). Shown is the distribution of mHG p-values obtained from 1,000 simulations of a long list (N=10,000), with K=500 and a small enrichment (1.5-fold) among the first 5,000 elements. The gray line indicates a significance threshold of $\alpha = 0.01$.}
        \label{fig:mhg_sens}
\end{figure}

For Scenario 1, imagine a relatively long list, (say, N=10,000), which has a very moderate enrichment (say, 1.5-fold) in the first half of the list. \cref{fig:mhg_sens} shows the distribution of mHG p-values obtained for 1,000 simulations of this scenario, for K=500. As can be seen from the distribution, the p-values obtained in these simulations are highly statistically significant. This is not surprising, since even a relatively small fold enrichment of 1.5 is extremely unlikely to arise by chance given a large enough sample. Therefore, a very good (i.e., small) mHG test statistic $\mhgstat = \min_n\{\hgp\}$ will be found at $n \approx 5,000$, and result in a highly significant $\mhgp$. However, in many applications, a slight overrepresentation of ``1's'' among the first half of the list may not represent a very interesting enrichment signal, since weak enrichment among a large part of the list could be artifactual, e.g. arising from a small and potentially unknown bias present in the data.


\begin{figure}[!ht]
        \centering
        \includegraphics[width=0.9\textwidth]{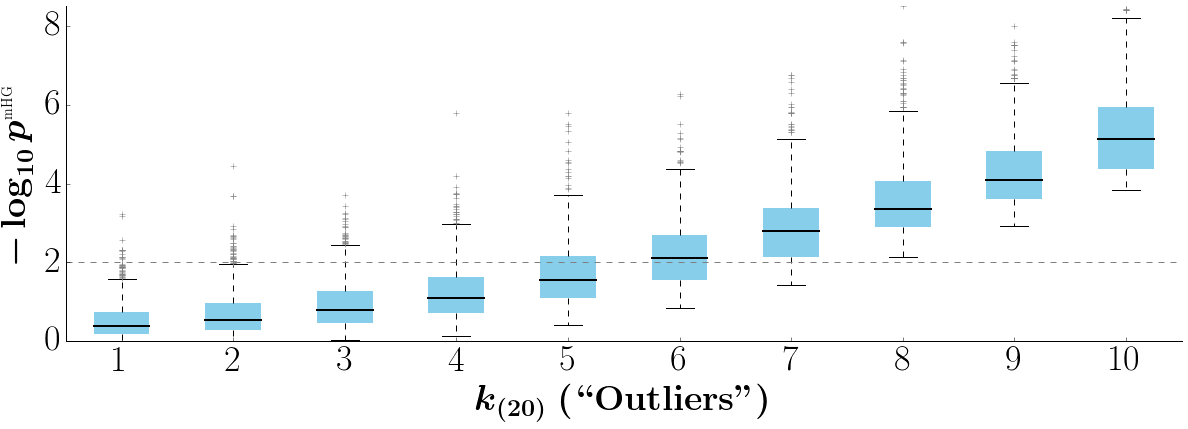}
        \caption{Assessing the robustness of the mHG to outliers (Scenario 2). Box plot showing the distributions of mHG p-values obtained from 1,000 simulations each for 1-10 ``outliers'' in lists with N=1,000 and K=100. The outliers are randomly distributed among the top 20 elements of the list, while the remaining 1's show no enrichment, i.e., they are randomly distributed across the entire list. The gray line indicates a significance threshold of $\alpha=0.01$.}
        \label{fig:mhg_robust}
\end{figure}

For Scenario 2, imagine a medium-sized list (say, N=1,000), with K=100 1's (i.e., ``interesting'' elements). Let us this time assume that there is no enrichment present at all (i.e., the 1's are randomly distributed), except for a few ``outliers'' at the top, which are randomly distributed among the first 20 positions in the list. How many of such outliers $k_{20}$ does it take for the mHG to yield a statistically significant $\mhgp$? \cref{fig:mhg_robust} shows boxplots for $k_{20}=1\dots{}10$, showing that for $k_{20}=6$, the majority of simulations result in a statistically significant $\mhgp$. Note that these positive test results are based on the high ranking of only 6/100 = 6\% of all the 1's in the list. It should be noted here that this extreme sensitivity can be thought of as a key feature of the mHG. However, this amazing sensitivity simultaneously makes the mHG vulnerable to outliers. One way to address this problem would be to perform a manual ``quality check'' on positive test results that are based on only very few 1's at the top of the list. An alternative strategy, which is presented in this work, is to introduce an additional parameter that directly controls the tradeoff between the test's sensitivity and robustness. 


\subsection{The XL-mHG test statistic}\label{sec:xlmhg_stat}

In both of the scenarios described in \cref{sec:xlmhg_limit}, a better control over the cutoffs tested by the mHG could help overcome the limitations encountered:

\begin{itemize}
\item In Scenario 1, the testing of very low cutoffs (large $n$) resulted in a positive test even though the enrichment pattern might be considered artifactual. To avoid this situation, we might want to limit the cutoffs tested to the first $L$ ranks. For example, we might decide that the lowest cutoff at which we would expect to find meaningful enrichment corresponded to $N/4$. This would significantly reduce the probability of obtaining a significant test result simply because of weak enrichment affecting the top 50\% of the list.
\item In Scenario 2, the high ranking of only 6\% of the 1's (``outliers'') was sufficient to obtain a positive test result in the majority of cases, even though the remaining 94\% of 1's exhibited no enrichment at all. To improve the robustness of our test, we might decide to ignore all cutoffs that have less than $X$ 1's above them. In our example, had we required at least 15\% (X=15) of 1's to be above the cutoff, this would have prevented the six outliers from generating a positive test result.
\end{itemize}

We therefore introduce the XL-mHG test statistic $\xlmhgstat$, which is a modification of $\mhgstat$:
\begin{align}\label{eq:xlmhgstat}
\tag{XL-mHG test statistic}
\xlmhgstat =
\begin{cases}
\displaystyle \min_{\substack{\kn \ge X \\n \le L}} \hgp & \textrm{if $\kL \ge X$,} \\
\qquad \quad \, 1 & \textrm{otherwise}
\end{cases}
\end{align}
This statistic introduces two parameters, $X$ and $L$ ($0 \le L \le N$, $X \ge 0$), as proposed above, which provide a certain level of control over which cutoffs should be tested for enrichment\footnote{The $L$ parameter was already discussed by Eden et al., who referred to it as $n_{\textrm{max}}$, and noted that ``it is possible to devise appropriate bounds and algorithms for computing the accurate p-value'' for such a parameter \cite{eden_discovering_2007}. This is exactly what we are concerned with here.} All cutoffs with less than $X$ 1's above them, as well as all cutoffs below $L$, are ignored. If there are less than $X$ 1's above the lowest permissable cutoff $L$, we have no enrichment at all ($\xlmhgstat = 1$). We immediately observe that for X=0 and L=N, $\xlmhgstat$ reduces to $\mhgstat$. Therefore, the XL-mHG test is a \textit{generalization} of the mHG test.

Note that instead of $X$ (or in addition to it?), we could also choose to introduce a parameter $T$, which, in analogy to $L$, would simply result in all cutoffs \textit{above} $T$ being ignored. However, we deliberately decide against this possibility, since for most applications, specifying $X$ is much more intuitive than specifying $T$.\footnote{An additional possibility is the testing of only certain ``slices'' of cutoffs, e.g., ${\{T_1 \dots{} L_1\}}\, \cup\, {\{T_2 \dots{} L_2\}}\, \cup\, \dots{}$ ($T_i < L_i$, $L_i < L_{i+1}$). Calculating the corresponding test statistics and p-values would require modifications of the respective algorithms that are similar to those introduced here.}

To efficiently calculate $\xlmhgstat$, we introduce a modification of \cref{alg:mhgstat}, with changes highlighted in magenta:
\begin{algorithm}[H]
\caption{Calculate $\textcolor{magenta}{\xlmhgstat}$, in $\mathcal{O}(KN)$}
\label{alg:xlmhgstat}
\begin{algorithmic}[1]
\REQUIRE V=$\bm{v}$, N=$N$, K=$K$, \textcolor{magenta}{X=$X$}, \textcolor{magenta}{L=$L$}
\ENSURE s=$\textcolor{magenta}{\xlmhgstat}$
\STATE k $\leftarrow$ 0
\STATE s $\leftarrow$ 1.0
\STATE F $\leftarrow$ \cref{alg:fkn_recursive} (V, N, K) \COMMENT{calculate all $\fkn$}
\FOR{n = 0 to \textcolor{Magenta}{L-1}}
\IF[we hit a ``1'']{V[n] != 0}
\STATE $k \leftarrow$ k + 1
\IF{\textcolor{magenta}{k $\ge$ X}}
\STATE p $\leftarrow$ \cref{alg:hgp_from_fkn} (F[n+1], k, N, K, n+1) \COMMENT{calculate $\hgp$}
\STATE s $\leftarrow \min$(s, p)
\ENDIF
\ENDIF
\ENDFOR
\RETURN s
\end{algorithmic}
\end{algorithm}

\pagebreak
\subsection{The XL-mHG p-value}\label{sec:xlmhg_pval}

The XL-mHG p-value $\xlmhgp$ is defined analogously to the \ref{eq:mhg_pval}: Assume that we observe $\xlmhgstat$ for a ranked binary list $\bm{v}$. Let $\xlmhgstatnullrnd$ be a random variable representing the XL-mHG test statistic observed for a random permutation of $\bm{v}$. Then:
\begin{align}\label{eq:xlmhg_pval}
\tag{XL-mHG p-value}
\xlmhgp = \Pr(\xlmhgstatnullrnd \le \xlmhgstat)
\end{align}

The introduction of $X$ and $L$ results in the elimination of certain configurations from $\mathcal{R}$: those with less than $X$ 1's above the cutoff, and those with cutoffs greater than $L$. Let $\RXL$ be this restricted set. Then we can express $\RXL$ as follows:
\begin{align*}
\RXL = \{\mukw : \pkw \le \xlmhgstat,\: k \ge X,\: k+w \le L\}
\end{align*}
In other words, for a configuration $\mukw$ to be in $\RXL$, we do not only require its associated hypergeometric p-value $\pkw$ to be at least as good as $\xlmhgstat$, but $k$ and $w$ must also fall within the limits defined by $X$ and $L$.

\begin{figure}[!htb]
        \centering
        \includegraphics[width=0.9\textwidth]{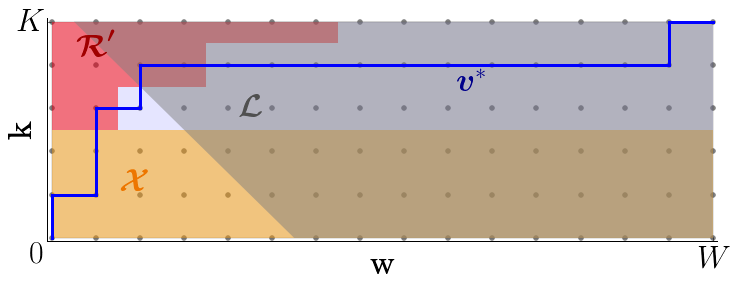}
        \caption{Expressing $\RXL$ using $\Rmod$, $\mathcal{X}$, and $\mathcal{L}$. The ${(K+1)}\times{(W+1)}$-grid shows all configurations in $\MKN$, as in \cref{fig:mhg_pval}. $\Rmod$ (red shaded area) is defined by $\xlmhgstat$ for the example vector $\vex$, with $X=3$ and $L=5$. $\mathcal{X}$, the set of all configurations excluded by $X$, is shown as the yellow shaded area. Similarly, $\mathcal{L}$, the set of all configurations excluded by $L$, is shown as the gray shaded area. $\RXL$ is the subset of $\Rmod$ contained in neither $\mathcal{X}$ nor $\mathcal{L}$.}
        \label{fig:xlmhg_pval}
\end{figure}

Alternatively, let us define $\Rmod = \{\mukw: \pkw \le \xlmhgstat\}$. We also define $\mathcal{X} = \{\mukw : k < X\}$ and $\mathcal{L} = \{\mukw : k + w > L\}$, representing the sets of configurations excluded by $X$ and $L$, respectively. We can then express $\RXL$ in terms of these sets (see \cref{fig:xlmhg_pval}):
\begin{align*}
\RXL = \Rmod \setminus (\mathcal{X} \cup \mathcal{L})
\end{align*}

\pagebreak
It should now be clear that we can easily modify \cref{alg:mhg_pval_R} to find all configurations in $\RXL$, as follows (changes highlighted in magenta):
\begin{algorithm}[H]
\caption{Determine whether $\mu_{(k,w)} \in \textcolor{magenta}{\RXL}$, for all $\mu_{(k,w)}$, in $\mathcal{O}(KW)$}
\label{alg:xlmhg_pval_R}
\begin{algorithmic}[1]
\REQUIRE s=$\textcolor{magenta}{\xlmhgstat} \in (0;1)$, K=$K$, W=$W$, \textcolor{magenta}{$X$=X}, \textcolor{magenta}{L=$L$}
\ENSURE Binary array R[0..K, 0..W], indicating whether $\mu_{(k,w)} \in \textcolor{magenta}{\RXL}$.
\STATE R $\leftarrow$ (K+1)x(W+1)-array of zeros
\STATE N $\leftarrow$ K+W
\STATE n $\leftarrow$ 1
\STATE p\_start $\leftarrow$ 1.0
\WHILE{\textcolor{magenta}{n $\le$ L}}
\STATE
\STATE // calculate $p_{(\knstar,n-\knstar)}$
\IF{n $\le$ K}
\STATE k $\leftarrow$ n
\STATE // calculate $\fn$ from $\fnmm$ using \ref{eq:ident4}
\STATE p\_start $\leftarrow$ p\_start * (K-n+1)/(N-n+1)
\ELSE
\STATE k $\leftarrow$ K
\STATE // calculate $\fK$ from $\fKmn$ using \ref{eq:ident5}
\STATE p\_start $\leftarrow$ p\_start * n/(n-K)
\ENDIF
\STATE
\STATE // find lowest $k$ for which $\mukw \in \mathcal{R}$
\STATE p $\leftarrow$ p\_start
\STATE pval $\leftarrow$ p\_start
\STATE w $\leftarrow$ n-k
\WHILE{\textcolor{magenta}{k $\ge$ X} and pval $\le$ s}
\STATE // we're still in $\RXL$
\STATE R[k,w] $\leftarrow$ 1
\STATE // calculate $\fkm$ from $\fk$ using \ref{eq:ident6}
\STATE p $\leftarrow$ p * (k*(N-K-n+k)) / ((n-k+1)(K-k+1))
\STATE pval $\leftarrow$ pval + p
\STATE k $\leftarrow$ k-1
\STATE w $\leftarrow$ w+1
\STATE
\ENDWHILE
\STATE n $\leftarrow$ n+1
\ENDWHILE
\RETURN R
\end{algorithmic}
\end{algorithm}

\pagebreak
Finally, using the return value of \cref{alg:xlmhg_pval_R}, we can use \cref{alg:mhg_pval} virtually unchanged to calculate $\xlmhgp$ (changes highlighted in magenta):

\begin{algorithm}[H]
\caption{Calculate $\textcolor{magenta}{\xlmhgp}$, in $\mathcal{O}(KW)$}
\label{alg:xlmhg_pval}
\begin{algorithmic}[1]
\REQUIRE s=$\textcolor{magenta}{\xlmhgstat} \in (0;1)$, K=$K$, W=$W$, \textcolor{magenta}{X=$X$}, \textcolor{magenta}{L=$L$}
\ENSURE p=$\textcolor{magenta}{\xlmhgp}$
\STATE \textcolor{magenta}{R $\leftarrow$ \cref{alg:xlmhg_pval_R} (s, K, W, X, L)}
\STATE M $\leftarrow$ (K+1)x(W+1)-array
\STATE M[0,0] $\leftarrow$ 1.0
\STATE N $\leftarrow$ K+W
\FOR{n = 1 to N}
\STATE k $\leftarrow$ min(n,K)
\STATE w = n-k
\WHILE{k $\ge$ 0 and w $\le$ W}
\IF{R[k,w] = 1}
\STATE M[k,w] $\leftarrow$ 0
\ELSIF{w $>$ 0 and k $>$ 0}
\STATE M[k,w] $\leftarrow$ M[k,w-1] * (W-w+1)/(N-n+1) + \\ \qquad M[k-1,w] * (K-k+1)/(N-n+1)
\ELSIF{w $>$ 0}
\STATE M[k,w] $\leftarrow$ M[k,w-1] * (W-w+1)/(N-n+1)
\ELSIF{k $>$ 0}
\STATE M[k,w] $\leftarrow$ M[k-1,w] * (K-k+1)/(N-n+1)
\ENDIF
\STATE w $\leftarrow$ w + 1
\STATE k $\leftarrow$ k - 1
\ENDWHILE
\ENDFOR
\STATE p $\leftarrow$ 1.0 - M[K,W]
\RETURN p
\end{algorithmic}
\end{algorithm}

\pagebreak

\section{Quantifying the strength of enrichment}\label{sec:enrich}

\subsection{Motivation}

\cref{sec:mhgimpl,sec:xlmhg} detail efficient algorithms to calculate the mHG and XL-mHG test statistics, respectively, as well as their associated p-values. When these quantities indicate the presence of significant enrichment, the next question becomes: How strong is the enrichment detected? In other words, how can we quantify the effect size of the enrichment, as opposed to its significance? In the case of the simple enrichment test that operated using a fixed cutoff (see \cref{sec:simple_enrich}), the answer to this question is easy: We can use the cutoff $n$ to calculate a \textit{fold enrichment} value $\esimple$, representing the ratio between the observed ($\kn$) and the expected ($\knnull$) number of 1's above the cutoff, where $\knnull$ is easily calculated as $K * (n/N)$:
\begin{align*}
\tag{fold enrichment}
\esimple = \frac{\kn}{\knnull} = \frac{\kn}{K * (n/N)}
\end{align*}

\begin{figure}[!ht]
        \centering
        \includegraphics[width=0.3\textwidth]{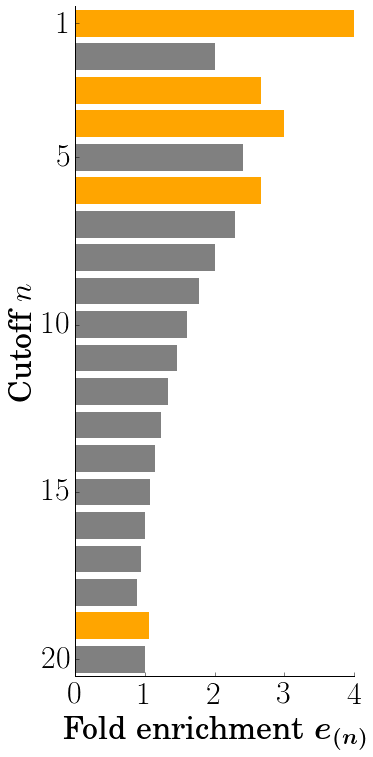}
        \caption{Quantifying the strenght of enrichment for the ranked binary list $\vex$. Each bar represents the fold enrichment $\esimple$ at a given cutoff $n$. Bars highlighted in yellow correspond to positions in the list that have a ``1'' element.}
        \label{fig:enrich}
\end{figure}

How can we adapt this simple definition to estimate the strength of enrichment in the absence of a fixed cutoff? It would at first seem natural to define a ``maximum fold enrichment'', in complete analogy to the \ref{eq:mhgstat}:
\begin{align*}
\tag{max. fold enrichment}
\emax = \max_n \esimple
\end{align*}
However, there is a clear problem with this approach: For small $n$, the behavior of $\esimple$ is very erratic. To demonstrate this, recall the example $\vex$:
\begin{align*}
\vex = (1,0,1,1,0,1,0,0,0,0,0,0,0,0,0,0,0,0,1,0)^T
\end{align*}
The values of $\esimple$ for this ranked list are shown in \cref{fig:enrich}. The largest fold enrichment is found for $n=1$, where $\esimple = 4.0$. In fact, this value is the largest fold enrichment value any list $\bm{v}$ with $N=20$ and $K=5$ can attain, since it corresponds to the situation where all of the elements above the cutoff consist of 1's. In particular, we have $\emax = 4.0$ for \textit{any} such ranked list which has a ``1'' as its first element. However, it would be rather useless to rely on a definition of enrichment that allows its value to be determined solely by the first element of the list. (Imagine this in a list with 1,000 elements!) Therefore, we should not use $\emax$ to quantify enrichment.


Since maximizing the fold enrichment over all cutoffs does not seem to work well, we could instead try to devise a way of selecting a ``special'' cutoff $\nstar$, and then use $\estar = \enstar$ as an overall measure of the strength of enrichment. A simple choice of $\nstar$ would be the the cutoff that determines the mHG test statistic:
\begin{align*}
\nstar = \argmin_{n}\, \hgp
\end{align*}
For $\vex$, $\nstar = 6$, and therefore $\estar = e_{(6)} \approx 2.7$, which seems like a much more reasonable value than $4.0$. In contrast to $\emax$, $\estar$ provides a useful estimate of the strength of enrichment that is no longer influenced by the fluctuations in $\esimple$ for small $n$.

However, the definition of $\estar$ is still somewhat unsatisfactory, in the sense that basing our estimate on the fold enrichment value at a single cutoff seems unjustified. After all, a significant mHG enrichment test result does not imply that $\nstar$ is the \textit{only} cutoff above which an enrichment of 1's can be observed. In fact, in our example of $\vex$, we see that $e_{(4)} = 3.0$, which is greater than $e_{(6)}$! Assuming that we have already established the general significance of enrichment in $\vex$ using the mHG test (${\mhgp \approx 0.024}$), why should we ignore $e_{(4)}$ in quantifying the strength of the enrichment for $\vex$?

\subsection{The mHG enrichment score $\mhge$}

The preceding discussion motivates us to find some middle ground in defining an mHG enrichment score $\mhge$: We do not want to include \textit{all} cutoffs in the calculation of $\mhge$, since for small $n$, the fold enrichment values are unreliable. Neither do we want to base $\mhge$ on the fold enrichment at a \textit{single} cutoff, since this seems unnecessarily restrictive. Instead, we observe that these two choices represent the two opposite extremes in a more general framework: Let us define a p-value threshold $\psi \ge \mhgstat$, and let $\Cpsi$ represent the set of cutoffs that are associated with hypergeometric p-values ${\hgp \le \psi}$:
\begin{align*}
\Cpsi = \big\{\, n : \hgp \le \psi\, \big\}
\end{align*}
Then, let ${\mhgepsi}$ be defined as follows:
\begin{align*}
\tag{mHG enrichment score}
\mhgepsi = \max_{n\, \in\, \Cpsi} \esimple
\end{align*}

In order for the fold enrichment at a specific cutoff to be included in the calculation of $\mhge$, the cutoff needs to be associated with a certain minimum significance of hypergeometric enrichment $\psi$. In particular, we observe that for $\psi = 1.0$, we have $\mhgepsi \equiv \emax$, and for $\psi = \mhgstat$, we have $\mhgepsi \equiv \estar$. Generally speaking, smaller values of $\psi$ better protect $\mhgepsi$ against fluctuations in $\esimple$ associated with small $n$, but they are also more likely to result in an overly conservative estimate of enrichment. Thus, the choice of $\psi$ determines the trade-off between robustness and accuracy in quantifying the strength of enrichment.

\subsection{The XL-mHG enrichment score $\xlmhge$}

In analogy to the generalization from $\mhgstat$ to $\xlmhgstat$, we would like to adapt the preceding definition of $\mhge$ for use in conjunction with the XL-mHG test. For this purpose, we require $\psi \ge \xlmhgstat$, and define $\Cxlpsi$ in analogy to $\Cpsi$, but restricted to the cutoffs permitted by $X$ and $L$:
\begin{align*}
\Cxlpsi &= \big\{\, n : \kn \ge X,\, n \le L,\: \hgp \le \psi\, \big\}
\end{align*}
We then define $\xlmhgepsi$ in analogy to $\mhgepsi$:
\begin{align*}
\tag{XL-mHG enrichment score}
\xlmhgepsi = \max_{n\, \in\, \Cxlpsi} \esimple
\end{align*}

As for $\xlmhgstat$, for $X=0$ and $L=N$, we have $\xlmhge \equiv \mhge$. Also, since $\xlmhge$ is only of interest when $\xlmhgp$ is considered significant (at a given significance level $\alpha$), $\psi$ can theoretically be set to a permissive value like $0.05$, even if $\alpha$ was chosen very conservatively (e.g., as a result of Bonferroni correction). However, any value $\psi > \alpha \ge \xlmhgstat$ will lead to a less conservatively biased estimate of enrichment than $\estar$.

\pagebreak
\section{References}
\printbibliography[heading=none]


\section{Acknowledgements}

I would like to thank Dr. Sandeep Dave for his support. I would further like to thank \href{http://bioinfo.cs.technion.ac.il/people/zohar/}{Dr. Zohar Yakhini} for introducing me to the mHG, and for providing me with very helpful comments and corrections during the preparation of parts of this manuscript. However, I remain solely responsible for any errors or omissions.


\section{Copyright and License}

Copyright (c) 2015 Florian Wagner. This work is licensed under a \href{http://creativecommons.org/licenses/by-nc-sa/4.0/}{Creative Commons Attribution-NonCommercial-ShareAlike 4.0 International License}.

\pagebreak

\appendix
\addappheadtotoc
\appendixpage
\label{sec:appendix}

\pagebreak

\section{Recurrence relations for the hypergeometric PMF}\label{sec:ident}

In this section, we simply state several recurrence relations that are used in the mHG algorithm, while postponing the derivations to \cref{sec:derivate}.

\begin{itemize}
\item Calculate $\fkp$ from $\fk$:
\begin{multline}\label{eq:ident1}
\tag{Identity 1}
\fkp =\\ \fk \frac{(n-k)(K-k)}{(k+1)(N-K-n+k+1)}
\end{multline}
See \cref{sec:deriv_ident1} for the derviation.

\item Calculate $\fkpn$ from $\fk$:
\begin{multline}\label{eq:ident2}
\tag{Identity 2}
\fkpn =\\ \fk \frac{(n+1)(N-K-n+k)}{(N-n)(n-k+1)}
\end{multline}
See \cref{sec:deriv_ident2} for the derviation.

\item Calculate $\fkpp$ from $\fk$:
\begin{align}\label{eq:ident3}
\tag{Identity 3}
\fkpp = \fk \frac{(n+1)(K-k)}{(N-n)(k+1)}
\end{align}
See \cref{sec:deriv_ident3} for the derviation.

\item Calculate $\fn$ from $\fnmm$:
\begin{align}\label{eq:ident4}
\tag{Identity 4}
\fnmm = \fn \frac{K-n+1}{N-n+1}
\end{align}
This assumes $n \le K$. See \cref{sec:deriv_ident4} for the derviation.

\item Calculate $\fK$ from $\fKmn$:

\begin{align}\label{eq:ident5}
\tag{Identity 5}
\fK = \fKmn \frac{n}{n-K}
\end{align}
See \cref{sec:deriv_ident5} for the derviation.

\item Calculate $\fkm$ from $\fk$:

\begin{align}\label{eq:ident6}
\tag{Identity 6}
\fkm = \fk \frac{k(N-K-n+k)}{(n-k+1)(K-k+1)}
\end{align}
See \cref{sec:deriv_ident6} for the derviation.

\end{itemize}

\pagebreak

\section{Algorithms}

\subsection{Efficient calculation of the mHG test statistic $\mhgstat$}

\begin{algorithm}[H]
\caption{Calculate $\hgp$ from $\fkn$, in $\mathcal{O}(K)$}
\label{alg:hgp_from_fkn}
\begin{algorithmic}[1]
\REQUIRE f=$\fkn$, k=$\kn$, N=$N$, K=$K$, n=$n$
\ENSURE p=$\hgp$
\STATE p $\leftarrow$ f
\WHILE{k $< \min$(n, K)}
\STATE // calculate $\fkp$ from $\fk$ using \ref{eq:ident1}
\STATE f $\leftarrow$ f * (p((n-k)(K-k))/((k+1)(N-K-n+k+1)))
\STATE p $\leftarrow$ p + f
\STATE k $\leftarrow$ k + 1
\ENDWHILE
\RETURN p
\end{algorithmic}
\end{algorithm}

\begin{algorithm}[H]
\caption{Calculate $\fkn$ for all $n$, in $\mathcal{O}(N)$}
\label{alg:fkn_recursive}
\begin{algorithmic}[1]
\REQUIRE V=$\bm{v}$, N=$N$, K=$K$
\ENSURE F=$\fkn$ for all $n=0\dots{}N$
\STATE F[0] $\leftarrow$ 1.0
\STATE k $\leftarrow$ 0
\FOR{n = 0 to N-1}
\IF{V[n] = 0}
\STATE // calculate $\fkpn$ from $\fk$ using \ref{eq:ident2}
\STATE F[n+1] = F[n] * ((n+1)*(N-K-n+k)) / ((N-n)(n-k+1))
\ELSE
\STATE // calculate $\fkpp$ from $\fk$ using \ref{eq:ident3}
\STATE F[n+1] = F[n] * ((n+1)*(K-k)) / ((N-n)*(k+1))
\STATE k $\leftarrow$ k + 1
\ENDIF
\ENDFOR
\RETURN F
\end{algorithmic}
\end{algorithm}

\begin{algorithm}[H]
\caption{Calculate $\mhgstat$, in $\mathcal{O}(KN)$}
\label{alg:mhgstat}
\begin{algorithmic}[1]
\REQUIRE V=$\bm{v}$, N=$N$, K=$K$
\ENSURE s=$\mhgstat$
\STATE k $\leftarrow$ 0
\STATE s $\leftarrow$ 1.0
\STATE F $\leftarrow$ \cref{alg:fkn_recursive} (V, N, K) \COMMENT{calculate all $\fkn$}
\FOR{n = 0 to N-1}
\IF[we hit a ``1'']{V[n] != 0}
\STATE $k \leftarrow$ k + 1
\STATE p $\leftarrow$ \cref{alg:hgp_from_fkn} (F[n+1], k, N, K, n+1) \COMMENT{calculate $\hgp$}
\STATE s $\leftarrow \min$(s, p)
\ENDIF
\ENDFOR
\RETURN s
\end{algorithmic}
\end{algorithm}

\pagebreak
\subsection{Efficient calculation of the mHG p-value $\mhgp$}

I first describe the algorithm to efficiently determine, for given $K$, $W$, and $\mhgstat$, whether $\mukw \in \mathcal{R}$, for all $\mukw$. A hypergeometric configuration $\mukw$ is in $\mathcal{R}$ if the hypergeometric p-value $\pkw$ associated with it is at least as ``good'' (i.e., equal to or smaller than) the observed mHG test statistic $\mhgstat$ (see \cref{sec:mhg_pvalue}). Similarly to the approach chosen for calculating $\mhgstat$, we avoid calculating $\pkw$ ``from scratch'', and rely on recurrence relations instead.

\begin{figure}[!ht]
        \centering
        \includegraphics[width=0.8\textwidth]{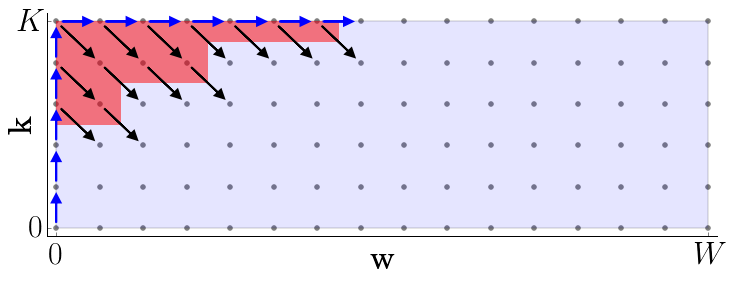}
        \caption{Illustration of \cref{alg:mhg_pval_R}. All hypergeometric configurations $\mukw$ are represented on an $(K+1)\times(W+1)$ grid, as in \cref{fig:mhg_pval}. The red shaded region contains all configurations that are in $\mathcal{R}$.}
        \label{fig:mhg_pval_R}
\end{figure}

Let $\knstar = \min\{n,K\}$. At each cutoff $n$, the algorithm first uses a recurrence relation to calculate the hypergeometric p-value $p_{(\knstar,n-\knstar)}$ for the configuration representing the strongest possible enrichment (see blue arrows in \cref{fig:mhg_pval_R}). If $p_{(\knstar,n-\knstar)} \in \mathcal{R}$, the p-value for the configuration with the next-lowest enrichment at cutoff $n$ is calculated using another recurrence relation, until a $p \notin \mathcal{R}$ is found (see black arrows in \cref{fig:mhg_pval_R}). The algorithm stops once it finds an $n$ for which $p_{(\knstar,n-\knstar)}$ is no longer in $\mathcal{R}$.

\begin{algorithm}[H]
\caption{Determine whether $\mu_{(k,w)} \in \mathcal{R}$, for all $\mu_{(k,w)}$, in $\mathcal{O}(KW)$}
\label{alg:mhg_pval_R}
\begin{algorithmic}[1]
\REQUIRE s=$\mhgstat \in (0;1)$, K=$K$, W=$W$
\ENSURE Binary array R[0..K, 0..W], indicating whether $\mu_{(k,w)} \in \mathcal{R}$.
\STATE R $\leftarrow$ (K+1)x(W+1)-array of zeros
\STATE N $\leftarrow$ K+W
\STATE n $\leftarrow$ 1
\STATE p\_start $\leftarrow$ 1.0
\WHILE{n $\le$ N}
\STATE
\STATE // calculate $p_{(\knstar,n-\knstar)}$
\IF{n $\le$ K}
\STATE k $\leftarrow$ n
\STATE // calculate $\fn$ from $\fnmm$ using \ref{eq:ident4}
\STATE p\_start $\leftarrow$ p\_start * (K-n+1)/(N-n+1)
\ELSE
\STATE k $\leftarrow$ K
\STATE // calculate $\fK$ from $\fKmn$ using \ref{eq:ident5}
\STATE p\_start $\leftarrow$ p\_start * n/(n-K)
\ENDIF
\STATE
\STATE // find lowest $k$ for which $\mukw \in \mathcal{R}$
\STATE p $\leftarrow$ p\_start
\STATE pval $\leftarrow$ p\_start
\STATE w $\leftarrow$ n-k
\WHILE{pval $\le$ s}
\STATE // we're still in $\mathcal{R}$
\STATE R[k,w] $\leftarrow$ 1
\STATE // calculate $\fkm$ from $\fk$ using \ref{eq:ident6}
\STATE p $\leftarrow$ p * (k*(N-K-n+k)) / ((n-k+1)(K-k+1))
\STATE pval $\leftarrow$ pval + p
\STATE k $\leftarrow$ k-1
\STATE w $\leftarrow$ w+1
\STATE
\ENDWHILE
\STATE n $\leftarrow$ n+1
\ENDWHILE
\RETURN R
\end{algorithmic}
\end{algorithm}

\pagebreak
The final algorithm for calculating $\mhgp$ relies on \cref{alg:mhg_pval_R} to determine $\mathcal{R}$, and then determines the number of paths that do not cross $\mathcal{R}$ using a simple recurrence relation (see \cref{sec:mhg_pvalue}).

\begin{algorithm}[H]
\caption{Calculate $\mhgp$, in $\mathcal{O}(KW)$}
\label{alg:mhg_pval}
\begin{algorithmic}[1]
\REQUIRE s=$\mhgstat \in (0;1)$, K=$K$, W=$W$
\ENSURE p=$\mhgp$
\STATE R $\leftarrow$ \cref{alg:mhg_pval_R} (s, K, W)
\STATE M $\leftarrow$ (K+1)x(W+1)-array
\STATE M[0,0] $\leftarrow$ 1.0
\STATE N $\leftarrow$ K+W
\FOR{n = 1 to N}
\STATE k $\leftarrow$ min(n,K)
\STATE w = n-k
\WHILE{k $\ge$ 0 and w $\le$ W}
\IF{R[k,w] = 1}
\STATE M[k,w] $\leftarrow$ 0
\ELSIF{w $>$ 0 and k $>$ 0}
\STATE M[k,w] $\leftarrow$ M[k,w-1] * (W-w+1)/(N-n+1) + \\ \qquad M[k-1,w] * (K-k+1)/(N-n+1)
\ELSIF{w $>$ 0}
\STATE M[k,w] $\leftarrow$ M[k,w-1] * (W-w+1)/(N-n+1)
\ELSIF{k $>$ 0}
\STATE M[k,w] $\leftarrow$ M[k-1,w] * (K-k+1)/(N-n+1)
\ENDIF
\STATE w $\leftarrow$ w + 1
\STATE k $\leftarrow$ k - 1
\ENDWHILE
\ENDFOR
\STATE p $\leftarrow$ 1.0 - M[K,W]
\RETURN p
\end{algorithmic}
\end{algorithm}

\pagebreak

\section{Derivations}\label{sec:derivate}

In these derivations, we omit terms that immediately cancel out because they appear identically in both enumerator and denominator.

\subsection{Derivation of \ref{eq:ident1}}\label{sec:deriv_ident1}

Using the definition of the \ref{eq:hypergeom}, we have:

\begin{align}\label{eq:hypergeom1}
\fkp = \frac{\binom{K}{k+1}\binom{N-K}{n-k-1}}{\binom{N}{n}}
\end{align}
Likewise, we have:

\begin{align}\label{eq:hypergeom2}
\binom{N}{n} = \frac{\binom{K}{k}\binom{N-K}{n-k}}{\fk}
\end{align}
By substituting $\binom{N}{n}$ in \eqref{eq:hypergeom1} with \eqref{eq:hypergeom2}, we then have:

\begin{align}
\MoveEqLeft[2] \fkp = \fk \frac{\binom{K}{k+1}\binom{N-K}{n-k-1}}{\binom{N-K}{n-k}\binom{K}{k}} \nonumber \\
&= \fkdot\frac{(n-k)!(N-K-n+k)!k!(K-k)!}{(k+1)!(K-k-1)!(n-k-1)!(N-K-n+k+1)!} \nonumber \\
&= \uuline{\fk\frac{(n-k)(K-k)}{(k+1)(N-K-n+k+1)}}
\end{align}

\subsection{Derivation of \ref{eq:ident2}}\label{sec:deriv_ident2}

Using the definition of the \ref{eq:hypergeom}, we have:

\begin{align}\label{eq:hypergeom3}
\fkpn = \frac{\binom{K}{k}\binom{N-K}{n-k+1}}{\binom{N}{n+1}}
\end{align}
Likewise, we have:

\begin{align}\label{eq:hypergeom4}
\binom{K}{k} = \fk \frac{\binom{N}{n}}{\binom{N-K}{n-k}}
\end{align}
By substituting $\binom{K}{k}$ in \eqref{eq:hypergeom3} with \eqref{eq:hypergeom4}, we then have:

\begin{align}
\MoveEqLeft[2] \fkpn = \fk \frac{\binom{N}{n}\binom{N-K}{n-k+1}}{\binom{N}{n+1}\binom{N-K}{n-k}} \nonumber \\
&= \fkdot\frac{(n+1)!(N-n-1)!(n-k)!(N-K-n+k)!}{n!(N-n)!(n-k+1)!(N-K-n+k-1)!} \nonumber \\
&= \uuline{\fk\frac{(n+1)(N-K-n+k)}{(N-n)(n-k+1)}}
\end{align}

\subsection{Derivation of \ref{eq:ident3}}\label{sec:deriv_ident3}

Using the definition of the \ref{eq:hypergeom}, we have:

\begin{align}\label{eq:hypergeom5}
\fkpp = \frac{\binom{K}{k+1}\binom{N-K}{n-k}}{\binom{N}{n+1}}
\end{align}
Likewise, we have:

\begin{align}\label{eq:hypergeom6}
\binom{N-K}{n-k} = \fk \frac{\binom{N}{n}}{\binom{K}{k}}
\end{align}
By substituting $\binom{N-K}{n-k}$ in \eqref{eq:hypergeom5} with \eqref{eq:hypergeom6}, we then have:

\begin{align}
\MoveEqLeft[2] \fkpp = \fk \frac{\binom{N}{n}\binom{K}{k+1}}{\binom{N}{n+1}\binom{K}{k}} \nonumber \\
&= \fkdot\frac{(n+1)! (N-n-1)! k! (K-k)!}{n! (N-n)! (k+1)! (K-k-1)!} \nonumber \\
&= \uuline{\fk\frac{(n+1)(K-k)}{(N-n)(k+1)}}
\end{align}

\subsection{Derivation of \ref{eq:ident4}}\label{sec:deriv_ident4}

We first derive the more general relation between $\fk$ and $\fkmm$, and then substitute $k=n$. By definition of the \ref{eq:hypergeom}, we have:

\begin{align}\label{eq:hypergeom7}
\fk = \frac{\binom{K}{k}\binom{N-K}{n-k}}{\binom{N}{n}}
\end{align}
Likewise, we have:

\begin{align}\label{eq:hypergeom8}
\binom{N-K}{n-k} = \fkmm \frac{\binom{N}{n-1}}{\binom{K}{k-1}}
\end{align}
By substituting $\binom{N-K}{n-k}$ in \eqref{eq:hypergeom7} with \eqref{eq:hypergeom8}, we then have:

\begin{align}
\MoveEqLeft[2] \fk = \fkmm \frac{\binom{K}{k}\binom{N}{n-1}}{\binom{N}{n}\binom{K}{K-1}} \nonumber \\
    &= \fkmm \frac{n! (N-n)! (k-1)! (K-k+1)!}{k! (K-k)! (n-1)! (N-n+1)!} \nonumber \\
    &= \fkmm \frac{n (K-k+1)}{k (N-n+1)}
\end{align}
Then, substituting $k=n$ (assuming $n \le K$):
\begin{align}
\fn &= \uuline{\fnmm \frac{K-n+1}{N-n+1}}
\end{align}

\subsection{Derivation of \ref{eq:ident5}}\label{sec:deriv_ident5}

We first derive the more general relation between $\fk$ and $\fkmn$, and then substitute $k=K$. By definition of the \ref{eq:hypergeom}, we have:

\begin{align}\label{eq:hypergeom9}
\fk = \frac{\binom{K}{k}\binom{N-K}{n-k}}{\binom{N}{n}}
\end{align}
Likewise, we have:
\begin{align}\label{eq:hypergeom10}
\binom{K}{k} = \fkmn \frac{\binom{N}{n-1}}{\binom{N-K}{n-k-1}}
\end{align}
By substituting $\binom{K}{k}$ in \eqref{eq:hypergeom9} with \eqref{eq:hypergeom10}, we then have:

\begin{align}
\MoveEqLeft[2] \fk = \fkmn \frac{\binom{N}{n-1}\binom{N-K}{n-k}}{\binom{N}{n}\binom{N-K}{n-k-1}} \nonumber \\
    &= \fkdot \frac{n! (N-n)! (n-k-1)! (N-K-n+k+1)!}{(n-1)! (N-n+1)! (n-k)! (N-K-n+k)!} \nonumber \\
    &= \fkdot \frac{n (N-K-n+k+1)}{(N-n+1)(n-k)}
\end{align}
Then, substituting $k=K$:
\begin{align}
\fK &= \uuline{\fKmn \frac{n}{n-K}}
\end{align}

\subsection{Derivation of \ref{eq:ident6}}\label{sec:deriv_ident6}

Using the definition of the \ref{eq:hypergeom}, we have:

\begin{align}\label{eq:hypergeom11}
\fkm = \frac{\binom{K}{k-1}\binom{N-K}{n-k+1}}{\binom{N}{n}}
\end{align}
Likewise, we have:

\begin{align}\label{eq:hypergeom12}
\binom{N}{n} = \frac{\binom{K}{k}\binom{N-K}{n-k}}{\fk}
\end{align}
By substituting $\binom{N}{n}$ in \eqref{eq:hypergeom11} with \eqref{eq:hypergeom12}, we then have:

\begin{align}
\MoveEqLeft[2] \fkm = \fk \frac{\binom{N-K}{n-k+1}\binom{K}{k-1}}{\binom{K}{k}\binom{N-K}{n-k}} \nonumber \\
    &= \fkdot \frac{k! (K-k)! (n-k)! (N-K-n+k)!}{(n-k+1)! (N-K-n+k-1)! (K-k+1)!} \nonumber \\
    &= \uuline{\fk \frac{k (N-K-n+k)}{(n-k+1)(K-k+1)}}
\end{align}

\end{document}